\documentclass[aps,twocolumn,showpacs]{revtex4}
\usepackage{amsmath}
\usepackage{amssymb}
\usepackage[final]{graphics}

\newcommand{\half} {\frac{1}{2}}
\newcommand{\e} { {\rm e} }
\newcommand{\dv} { {\rm d} }
\newcommand{\iv} { {\it v} \phi_b^2 }
\newcommand{\kma} { k_m^2 a^2}

\newcommand{\yy} {\zeta}
\newcommand {\ys} {\left|\yy_s\right|}

\newcommand {\cs} {c_{\rm salt} }
\newcommand {\yfc} {\yy_c}
\newcommand {\pfc} {\eta_c}
\newcommand {\xfc} {x_c}

\newcommand {\cm} {\phi_M^2}
\newcommand {\cmz} {\phi_{M,0}^2}

\begin{document}

\title{ Polyelectrolytes Adsorption:
    Chemical and Electrostatic Interactions\\}

\author {Adi Shafir}
\email{shafira@post.tau.ac.il}
\affiliation{School of Physics and Astronomy   \\
    Raymond and Beverly Sackler Faculty of Exact Sciences \\
    Tel Aviv University, Ramat Aviv, Tel Aviv 69978, Israel\\}

\author{David Andelman}
\email{andelman@post.tau.ac.il}
\affiliation{School of Physics and Astronomy   \\
    Raymond and Beverly Sackler Faculty of Exact Sciences \\
    Tel Aviv University, Ramat Aviv, Tel Aviv 69978, Israel\\}

\bigskip
\date{August 22, 2004}
\bigskip
\vskip 1truecm
\begin{abstract}
\setlength{\baselineskip} {14pt} \linespread{1.6}

Mean-field theory is used to model polyelectrolyte adsorption and
the possibility of overcompensation of charged surfaces. For
charged surfaces that are also chemically attractive, the
overcharging is large in high salt conditions, amounting to
$20-40\%$ of the bare surface charge. However, full charge
inversion is not obtained in thermodynamical equilibrium for physical
values of the parameters. The overcharging increases with
addition of salt, but does not have a simple scaling form with the
bare surface charge. Our results indicate that more evolved
explanation is needed in order to understand polyelectrolyte
multilayer built-up. For strong polymer-repulsive surfaces, we
derive simple scaling laws for the polyelectrolyte adsorption and overcharging.
We show that the overcharging scales linearly with the bare
surface charge, but its magnitude is very small in comparison to
the surface charge. In contrast with the attractive surface, here
the overcharging is found to decrease substantially with addition
of salt. In the intermediate range of weak repulsive surfaces, the
behavior with addition of salt crosses over from increasing
overcharging (at low ionic strength) to decreasing one (at high
ionic strength).  Our results for all types of surfaces are
supported by full numerical solutions of the mean-field equations.

\end{abstract}
\pacs{82.35.Gh, 82.35.Rs, 61.41.+e}

\maketitle
\newpage
\section{Introduction}
\label{Intro} \setlength{\baselineskip} {14pt} \linespread{1.6}

Aqueous solutions containing  polyelectrolytes and small ions are
abundant in biological systems, and have been the subject of
extensive research in recent years. When such a solution is in
contact with an oppositely charged surface, adsorption of the
polyelectrolyte chains can occur. Theoretical descriptions of
polyelectrolyte adsorption take into account the multitude of
different interactions and length scales. Among others they
include electrostatic interactions between the surface, monomers
and salt ions, excluded-volume interactions between monomers and
entropy considerations. Although a full description of
polyelectrolytes is still lacking at present, several approaches
exist and use different types of approximations
\cite{wiegel}-\cite{muthu3}. These include linearized mean-field
equations \cite{joanny,chatellier,varoqui1,varoqui2,muthu,wiegel},
numerical solutions of non-linear the mean-field equations
\cite{itamar1,itamar2,itamar3,us}, scaling considerations
\cite{itamar1,itamar2,itamar3,us,dobrynin,borisov,manghi,netz,review,review1},
multi-Stern layers of discrete lattice models
\cite{fleer,vanderschee,vandesteeg,bohmer} and computer
simulations \cite{yamakov,muthu2,muthu3}.

Experimental studies \cite{mohwald,decher,regine} have shown that
adsorbing polyelectrolytes (PEs) may carry a charge greater than
that of the bare surface, so that the overall
surface-polyelectrolyte complex has a charge opposite to that of
the bare charged surface. This phenomenon is known as {\it
overcharging} (or surface charge {\it overcompensation}) by the PE
chains. When the overcharging is large enough to completely
reverse the bare surface charge, the resulting charge surplus of
the complex can be used to attract a second type of
polyelectrolyte having an opposite charge to that of the first
polyelectrolyte layer. Eventually, by repeating this process, a
complex structure of alternating layers of positively and
negatively charged polyelectrolytes can be formed. Experimentally,
multilayers consisting of hundreds of such layers can be created
\cite{mohwald,decher}, leading the way to several interesting
applications.

A theoretical description of the PE overcharging was proposed in
Refs.~\cite{joanny,castelnovo,castelnovo2} based on a mean-field
formalism. The model relies on several approximations for a very dilute
PE solution in contact with a charged surface and in theta solvent condition.
In the high salt limit, the model
predicts an exact charge inversion for {\it indifferent} ({\it i.e.}
non-interacting) surfaces.
In another work~\cite{itamar1}, the
scaling of the adsorption parameters was derived using a
Flory-like free energy. This work used mean field theory, but with a different type
of boundary conditions.  Using
a strongly (non-electrostatic) repulsive surface for the PE, profiles of monomer
concentration and electrostatic potential have been calculated
from numerical solution of the mean-field equations. Several scaling laws have been proposed
and the possibility of
a weak overcharging in low salt condition has been demonstrated.
In a related work~\cite{us} we
presented simple scaling laws resulting from the same mean-field
equations and for the same chemically repulsive surfaces. In particular,
we addressed
the adsorption-depletion crossover as function of added salt. We showed that
addition of salt eventually causes the polyelectrolyte to deplete
from the charged surface, and pre-empts the high-salt adsorption
regime described in Ref.~\cite{itamar1}.

The present article can be regarded as a sequel of Ref.~\cite{us},
offering a more complete treatment of the adsorption problem for
several types of charged surfaces. In particular, it includes the
effect of surface-PE chemical interactions, and the scaling of
overcharging and adsorption in presence of added salt. These
chemical interactions are found  to play a crucial role in the
overcharging, showing that a necessary condition for the formation
of multilayers is a chemical attraction of non-electrostatic
origin (complexation) between the two types of PE chains as well
as between the PE and the charged surface. For chemically
attractive surfaces our results deviate from those of
Refs.~\cite{joanny,castelnovo,castelnovo2}. We find in the same
solvent and surface conditions that the overcharging does not
reach a full 100\% charge inversion of the bare surface charge. It
rather depends on system parameters and never exceeds 30-40\% for
physical range of parameters. Like in Ref.~\cite{joanny}, we find
an increase of the overcharging with salt but our numerical
results do not agree with the previous prediction. For repulsive
surfaces, several scaling laws are obtained and agree well with
the numerically obtained profiles. However, due to the competition
between the electrostatic and chemical surface interactions, the
overcharging here is usually quite small, of the order of $\sim
1\%$ only.

The article is organized as follows: the mean-field equations,
used in the past in several other works for PE adsorption, are
reviewed in Sec.~\ref{mfeq}. The following two sections treat two
different types of surfaces. In Sec.~\ref{phimax}, results for PE
adsorption and overcharging for attractive surfaces are presented
and compared with previous models. In Sec.~\ref{ocreg} we derive
the scaling of adsorbed PE layers in the case of a chemically
repulsive surface.  In Sec.~\ref{append2}
we present the adsorption and overcharging for the intermediate case of
a weakly repulsive surface. In particular, we show the dependence of PE
adsorption on the solution ionic strength.  A summary of the main
results and future prospects are presented in
Sec~\ref{discussion}.

\section{The Mean-Field Equations and Surface Characteristics}
\label{mfeq}

Consider an aqueous solution, containing a bulk concentration of
infinitely long polyelectrolyte (PE) chains, together with their
counterions and a bulk concentration of salt ions. Throughout this
paper we assume for simplicity that the PE are positively charged and that the counterions
and added salt are all monovalent.
The mean-field equations describing such an ionic
solution have been derived elsewhere~\cite{us,itamar1} and are
briefly reviewed here:

\begin{equation}
  \nabla^2\psi= \frac{8\pi e \cs}{\varepsilon} \sinh {\beta e\psi} +
    \frac {4\pi e}{\varepsilon}\left(\phi_b^2 f\e^{\beta e\psi}-
    f\phi^2\right)
 \label{PoissonBoltzmanmod}
\end{equation}
\begin{equation}
  \frac{a^2}{6}\nabla^2\phi={\it{v}}\left(\phi^3-\phi_b^2\phi\right)+
    \beta f e\psi\phi ~,
 \label{Edwardsmod}
\end{equation}
where the polymer order parameter, $\phi({\bf r})$, is the square
root of $c({\bf r})$, the local monomer concentration, $\phi_b^2$
the bulk monomer concentration, $\psi$ the local electrostatic
potential, $\cs$ the bulk salt concentration, $\varepsilon$ the
dielectric constant of the aqueous solution, $f$ the monomer
charged fraction, $e$ the electron charge, $\it{v}$ the second
virial (excluded volume) coefficient of the monomers, $a$ the
monomer size and $\beta=1/k_BT$ the inverse of the thermal energy.
Eq.~(\ref{PoissonBoltzmanmod}) is the Poisson-Boltzmann equation
where the salt ions, counterions and monomers are regarded as
sources of the electrostatic potential. Eq.~(\ref{Edwardsmod}) is
the mean-field (Edwards) equation for the polymer order parameter
$\phi({\bf r})$, where the excluded-volume interaction between
monomers and external electrostatic potential $\psi({\bf r})$ are
taken into account.

For the case of an infinite planar wall at $x=0$,
Eqs.~(\ref{PoissonBoltzmanmod}) and (\ref{Edwardsmod}) can be
transformed into two coupled ordinary differential equations,
which depend only on the distance $x$ from the surface.

\begin{equation}
  \frac{\dv^2\yy}{\dv x^2}=\kappa^2\sinh \yy +k_m^2 \left(\e^\yy-
    \eta^2 \right)\\
 \label{normPBmod}
\end{equation}
\begin{equation}
  \frac{a^2}{6}\frac{\dv ^2\eta}{\dv x^2}=\iv
    \left(\eta^3-\eta \right)+f \yy \eta ~,
 \label{normEdwardsmod}
\end{equation}
where $\yy\equiv e\psi/k_BT$ is the dimensionless (rescaled)
electrostatic potential, $\eta^2\equiv\phi^2/\phi_b^2$ the
dimensionless monomer concentration, $\kappa^{-1}=\left(8\pi l_B
\cs\right)^{-1/2}$  the Debye-H\"{u}ckel screening length,
 due to added
salt concentration, $k_m^{-1}=\left(4\pi l_B \phi_b^2
f\right)^{-1/2}$ a similar decay length due to
counterions, and $l_B=e^2/\varepsilon k_BT$ the Bjerrum length.
For water with dielectric constant $\varepsilon=80$ at room
temperature, $l_{\rm B}$ is equal to about $7$\AA. Note that the
actual decay of the electrostatic potential is determined by a
combination of salt, counterions, and polymer screening effects.

The solution of Eqs.~(\ref{normPBmod}) and (\ref{normEdwardsmod})
requires four boundary conditions for the two profiles. The electrostatic
potential decays to zero at infinity, $\yy(x\rightarrow\infty)=0$. At the
surface, we have chosen to work with a constant surface potential. Namely, a
conducting surface with a potential $\psi=\psi_s$, or in rescaled variables,
$\yy(0)=-\ys$. The results can be easily extended to the case of fixed
surface charge density (which amounts to fixing the derivative of the
potential $\left.\dv\yy/\dv x\right|_{x=0}=-4\pi l_B\sigma$).

For the PE profile we also have two boundary conditions. At infinity,
$\eta(x\rightarrow\infty)=1$, because $\phi(\infty)=\phi_b$, has to match the
bulk value. The special case of a zero bulk monomer concentration can be
treated by taking $\phi_b\to 0$ and working directly with the non-rescaled
concentration, $\phi(x)$.

We model separately two types of surfaces. Although the surface always
attracts the oppositely charged PE chains, it can be either chemically
repulsive or attractive. In the case of a  chemical
repulsion between the PE chains and the surface, the amount
of PE chains in direct contact with the surface is set to a small
value $\eta(x=0)=\eta_s$. In the limit of a strong repulsive surface,
$\eta_s$ tends to zero and the boundary condition is $\eta(x=0)=0$.
Because of the ever-present longer-range electrostatic attraction with the
surface, PE chains accumulate in the surface vicinity,
resulting in a positive slope  $\left.\dv\eta/\dv x\right|_{x=0}>0$.

For chemical attractive surfaces we rely on a boundary condition often used
\cite{joanny,degennes2}
for neutral polymer chains: $\dv\eta(0)/\dv
x+ \eta(0)/d=0$ where $d$ has units of length and is inversely
proportional to the strength of non-electrostatic interactions of
the PEs with the surface. The limit of $d\to \infty$ is the {\it indifferent}
(non-interacting) surface limit, while the previous limit of a strong
repulsive surface is obtained by a small and negative $d$.

In the following sections we present our results, first for the chemically
attractive surface and then for the chemically repulsive one.

\section{Overcharging of Chemically Attractive Surfaces}
\label{phimax}

In this section we restrict the attention to surfaces that attract
the polymer in a non-electrostatic (short range) fashion. For
example, a possible realization can be a system containing PE chains
with hydrophobic groups (like polystyrene sulfonate) that are attracted
to a hydrophobic surface, like the water-air interface. The short range
attraction is modeled by a surface interaction in a similar way that
is used extensively for neutral polymers via a boundary condition on the
polymer order parameter at the surface $x=0$:

\begin{equation}
\eta(0)=\left.-d\frac{\dv\eta}{\dv x}\right|_{x=0}
 \label{nonelecbc}
\end{equation}
where the length $d$ was introduced in Sec.~II and is inversely
proportional to the strength of the surface interaction.

Close to a surface the polymer has a concentration profile given
in terms of the distance from the surface $x$. Within mean-field
theory the adsorbed polymer amount is defined with respect to the
bulk concentration $\phi^2_b$ to be:

\begin{equation}
  \Gamma\equiv \int_0^\infty\dv x\left(\phi^2-\phi^2_b\right)=
  \phi_b^2\int_0^\infty\dv x \left(\eta^2-1\right),
 \label{gammaDdef}
\end{equation}
We note that the definition of $\Gamma$ manifests one of the
deficiencies of mean-field theory where it is not possible to
distinguish between chains absorbed on the surface and those
accumulated in the surface vicinity. The latter will be washed
away when the surface is placed in a clean aqueous solution and
does not participate in the effective PE surface build-up.

Another important quantity to be used throughout this paper is the
overcharging parameter defined as the excess of PE charge over
that of the bare surface charge per unit area, $\sigma$:
\begin{equation}
  \Delta\sigma\equiv  -|\sigma|+ f\Gamma\, ,
 \label{delsig0}
\end{equation}
where $\sigma<0$ is the {\it induced} surface charge density
calculated from the fixed surface potential ($\yy^\prime=-4\pi l_B\sigma$)
in units of $e$, the electron charge. Note that in terms of the relative overcharging
parameter  $\Delta\sigma/|\sigma|$,
$-1\le \Delta\sigma/|\sigma| \le 0$ corresponds to {\it undercharging},
while a positive $\Delta\sigma/|\sigma| > 0$ to overcharging. The special value
$\Delta\sigma/|\sigma| =1$ indicates a {\it full} charge inversion.

We solved numerically the mean-field coupled equations,
Eqs.~(\ref{normPBmod})-(\ref{normEdwardsmod}) with the
electrostatic boundary condition of a fixed surface potential,
$\ys$ (that has a one-to-one correspondence with an induced
surface charge density, $\sigma$) and a non-electrostatic boundary
condition from Eq.~(\ref{nonelecbc}). The relative overcharging
$\Delta \sigma/|\sigma|$ is plotted in Fig.~1 as function of the
amount of salt. The solid line represents our numerical results,
while the dashed one corresponds to a previous prediction
\cite{joanny}. The same system parameters are used for both: the
limit of a dilute PE reservoir ($\phi_b=0$), theta solvent
conditions ($v=0$), a strong ionic strength, and an indifferent
surface taken in the limit of $d\to \infty$ (obtained already for
$d\ge 100$\,\AA). Figure 1 shows an increasing dependence of the
overcharging parameter on $\cs$, but does not obey any simple
scaling law. It varies on quite a large range of values from less
than 10\% (relative to $|\sigma|$) for $\cs\simeq 0.1$\,M to about
55\% for $\cs\simeq 5.5$\,M. We never observed for reasonable
values of the parameters a full charge inversion of 100\% or more.

Our results are contrasted with the approximated prediction of
Ref.~\cite{joanny}. In the high salt limit the prediction reads:

\begin{equation}
\frac{\Delta\sigma}{|\sigma|} = 1+\frac{2a^2}{3d f |\sigma|}\cs
\label{joannydelsig}
\end{equation}
where we note that in Ref.~\cite{joanny} all lengths are rescaled
with $a/\sqrt{6}$. This prediction is shown by the dashed line in
Fig.~1. One can see that the linear dependence on salt
concentration is very weak and that for the entire range of salt
the result is dominated by the first and constant term in
Eq.~(\ref{joannydelsig}), giving an overcharging of 100\%-110\%.
Although the general trend of an increased in
$\Delta\sigma/|\sigma|$ appears in both results, there is neither
agreement in the values of $\Delta\sigma/|\sigma|$ nor in the
quantitative dependence on $\cs$ for a wide range of $\cs$ values.

The results of Fig.~1 have been done in the limit of an
indifferent surface modeled by $d\gg\kappa^{-1}$. A closer
examination of the overcharging $d$-dependence is presented in
Fig.~2. Both in part (a) for $f=0.2$ and in part (b) for $f=1$,
our result and the prediction of Eq.~(\ref{joannydelsig}) show a
similar descending trend with $d$. Already for $d$ as small as
100\AA\ the asymptotic results for the indifferent surface result
is obtained. For smaller
$d$, the chemical attraction causes a bigger overcharging. The
main discrepancy between our exact solution of the mean-field equations
and Eq.~(\ref{joannydelsig}) is in the actual limiting value for
$d\to\infty$ and the lack of charge inversion from our results
(solid line in Fig.~2). Note also that the difference between the two
results cannot be explained by a constant multiplicative factor, and
that $\Delta\sigma/|\sigma|$ is smaller for $f=1$ than
for $f=0.2$ due to the larger electrostatic repulsion between
charged monomers.

To complete the presentation of the attractive surface we show in
Fig.~3 the dependence of the overcharging on $\cs$ for several
different values of surface potential, ranging from $\ys=1$ to
0.2. In part (a) the PE absorbed amount $\Gamma$  (in units of \AA$^{-2}$) is
shown while in part (b) the dimensionless $\Delta\sigma/|\sigma|$
is shown. The general trend is an increase of both quantities with
$\cs$ due to the non-electrostatic attraction of the PE chain
to the surface, on one hand, and the screening of the
monomer-monomer repulsion, on the other. The overcharging is less
than 100\% unless the amount of salt is unrealistically large. Another
observation can be seen from the behavior of
$\Delta\sigma/|\sigma|$. We clearly see that $\Delta\sigma$ does
not depend linearly on $\sigma$ since the three different $\ys$
give three different curves (no data collapse).

This very last result should be compared with the chemically
repulsive surfaces which is discussed next and for which we find
that $\Delta \sigma \sim \sigma$.

\section{Strong Chemically Repulsive Surfaces}
\label{ocreg}

The chemical
repulsion between the PE chains and the surface causes the amount
of PE chains in direct contact with the surface to diminish or
even be zero for the strong repulsive case.  The latter
limit is incorporated into the mean-field equations by taking the
boundary condition $\eta_s=0$, used previously in Refs.~\cite{itamar1,us}.

Our assumptions for treating the adsorbed layer are as follows. i)
Inside the adsorption layer, the electrostatic interactions are
assumed to be stronger than the excluded-volume interactions. ii)
Using results from Ref.~\cite{us}, we assume that the
electrostatic potential decays mainly via the PE adsorption and
not via the salt (for weak enough ionic strength). iii) Another
assumption is that  the electrostatic potential, $\yy\equiv
e\psi/k_BT$ can be written in terms of a scaling function $\yy=\ys
h(x/D)$, where $\ys$ is the surface potential and $D$ is the
adsorption layer length scale (see also Ref.~\cite{us}). iv) The
last assumption is that the electrostatic potential $\yy$ is low
enough so that we can employ the linear Debye-H\"uckel
approximation.

%

\subsection{First Integration of the Mean-Field Equations}
\label{app1}

The salt dependence of the adsorption characteristic can be
obtained from the first integration of Eqs.~(\ref{normPBmod})
and (\ref{normEdwardsmod}).  Multiplying Eq.~(\ref{normPBmod}) by
$\dv \yy / \dv x$ and Eq.~(\ref{normEdwardsmod}) by $\dv \eta /
\dv x$, and then integrating both equations from $x$ to infinity
yields:

\begin{eqnarray}
  \half\left(\frac{\dv \yy}{\dv x}\right)^2=\kappa^2\left(\cosh
  \yy-1\right)+
    k_m^2 \bigg(\e^{\yy}-1+ \nonumber \\
    \int_x^\infty\eta^2\frac{\dv \yy}{\dv x}\dv x\bigg)
 \label{PBfrstint}
\end{eqnarray}
\begin{equation}
  \frac{a^2}{12}\left(\frac{\dv \eta}{\dv x}\right)^2 =\frac{1}{4}\iv\left(
    \eta^2-1\right)^2-f\int_x^\infty  \eta \yy \frac{\dv \eta}{\dv x} \dv x.
 \label{EDfrstint}
\end{equation}
 Eq.~(\ref{EDfrstint}) is then multiplied by $2k_m^2/f$ and
 subtracted
from Eq.~(\ref{PBfrstint}). Using $\int_x^\infty (\dv \yy/\dv
x)\eta^2\,\dv x=-\eta^2\yy- 2\int_x^\infty (\dv \eta/\dv
x)\yy\eta\,\dv x$ we get:

\begin{eqnarray}
  \half\left(\frac{\dv \yy}{\dv x}\right)^2-
    \frac{\kma}{6f}\left(\frac{\dv \eta}
    {\dv x}\right)^2=\kappa^2\left(\cosh \yy-1\right)+  \nonumber
    \\
    k_m^2 \left(\e^\yy-1\right)- k_m^2\yy\eta^2-\frac{1}{2f}\iv k_m^2
    \left(\eta^2-1\right)^2, \label{PressureEq}
\end{eqnarray}
which can be interpreted as the local pressure balance equation.
The first term in the LHS of Eq.~(\ref{PressureEq}) is the
electrostatic field pressure, and the second term is the pressure
arising from chain elasticity. The first and second terms in the
RHS are the ideal gas pressure of the salt and counterions. The
third term is the pressure due to the interaction between the
electrostatic field and the monomer concentration, and the last
term is the excluded-volume driven pressure.

For every segment where $\yy$ is a monotonic function of $x$, a
change of variables from $x$ to $\yy$ can be performed. Using
${\dv \eta}/{\dv x} = {\dv \eta}/{\dv \yy}\cdot{\dv \yy}/{\dv x}$
in Eq.~(\ref{PressureEq}) yields:
\begin{eqnarray}
  \left(1-\frac{\kma}{3f}\left(\eta'(\yy)\right)^2\right)
    \left(\frac{\dv \yy}{\dv x}\right)^2=
    2\kappa^2\left(\cosh \yy -1\right)+ \nonumber \\
    2k_m^2 \left(\e^\yy -1-\yy\eta^2\right)-\frac{1}{f}\iv k_m^2
    \left(\eta^2-1\right)^2
 \label{modPressureEq}
\end{eqnarray}
Below, for the strong repulsive case, we attempt to produce approximate solutions to
Eqs.~(\ref{PressureEq}) and (\ref{modPressureEq}), and compare
them to numerical calculations of Eqs.~(\ref{normPBmod}) and
(\ref{normEdwardsmod}) (see also Ref.~\cite{us}).


Under the above assumptions, the excluded-volume term in Eq.~(\ref{modPressureEq})
can be neglected,
and $\exp(\yy)$
and $\cosh(\yy)$ can be expanded out to second order in $\yy$, yielding:

\begin{eqnarray}
  \left(\frac{\dv \yy}{\dv x}\right)^2-
  \frac{\kma}{3f}\left(\frac{\dv \eta}{\dv \yy}\right)^2
    \left(\frac{\dv \yy}{\dv x}\right)^2=\kappa^2\yy^2+ \nonumber
    \\
    2k_m^2\yy\left(1-\eta^2\right)+k_m^2\yy^2.
 \label{pressele}
\end{eqnarray}
The first term on the RHS relates to the salt ions, the second to
both the monomer and counterion concentrations and the third term to
the counterion concentration.

\subsection{Scaling of Potential and Concentration Profiles\\}
\label{scele}

Under the above assumptions, the dominant term in the RHS of
Eq.~(\ref{pressele}) is the second one related to the monomer and
counterion contributions. This term changes sign from negative to
positive at $\eta\simeq 1$. For $\eta\to 0$ (close to the
surface) the negative sign of the RHS implies that the second
(negative) term of the LHS is dominant: $\dv \eta / \dv \yy
>\sqrt{3f/\kma}$. Finally, it can be shown self-consistently
(presented below) that the first term on the LHS is of order
$\ys^3$ while the correction terms on the RHS are of order
$\ys^2$. Hence, close to the surface we neglect the first term in
the LHS.

\begin{eqnarray}
  \frac{\kma}{3f}\left(\frac{\dv \eta}{\dv \yy}\right)^2
    \left(\frac{\dv \yy}{\dv x}\right)^2=
    2k_m^2|\yy|\left(1-\eta^2\right) -  \nonumber\\
    \left(k_m^2+\kappa^2\right)\yy^2
 \label{pressele1}
\end{eqnarray}

On the charged surface $\eta=0$ and $\yy=-\ys$. Using the
mean-value theorem in the interval of interest $0\le\eta\le 1$ we
get $\left.\dv \eta / \dv \yy \right|_{y=y_s}\simeq 1/\ys$.
Substituting the scaling hypothesis $\left.\dv \yy /\dv
x\right|_{x=0}\simeq \ys / D$ in Eq.~(\ref{pressele1}) evaluated
at the surface yields an estimate for the layer thickness $D$:

\begin{eqnarray}
  D\simeq\frac{a}{\sqrt{6f\ys}}
  \left(1-\frac{\kappa^2+k_m^2}{2k_m^2}\ys\right)^{-1/2}
  \simeq  \nonumber \\
    \frac{a}{\sqrt{6f\ys}}\left(1+
    \frac{1}{4}\ys +\frac{\cs}{2f\phi_b^2}\ys\right).
 \label{Dele}
\end{eqnarray}
The first term of Eq.~(\ref{Dele}) retrieves the result given in
Refs.~\cite{us,itamar1,dobrynin,borisov}. The other terms include
corrections due to the ionic strength of the solution. As salt is
added, the monomer-monomer electrostatic repulsion wins over the
electrostatic attraction of the PEs to the surface, resulting in
an increase in the adsorption length $D$, and at very high salt
in depletion\cite{us}.


By changing $x$ to the dimensionless length $x/D$, neglecting the
excluded-volume term and inserting the potential scaling
hypothesis in Eq.~(\ref{normEdwardsmod}), the scaling form for the
monomer concentration $\phi_b^2\eta^2$ can be derived. The scaling
form of $\zeta$ (see assumption iii above) dictates a similar
scaling form: $\eta=\phi(x)/\phi_b\simeq
\sqrt{\cm/\phi_b^2}\,g(x/D)$, where $\cm$ is the peak monomer
concentration, and $g$ is a scaling function normalized to one at
the peak and satisfies $g(0)=0$.

To find $\cm$, the peak concentration condition $\dv \eta/\dv \yy
=0$ can be inserted in Eq.~(\ref{pressele}), causing the second
term in the LHS to vanish. Using the scaling for $\yy$ and $D$ of
Eq.~(\ref{Dele}) yields:

\begin{equation}
  \cm=\left(\phi_b^2+\frac{3\ys^2}{4\pi l_B a^2}\right)\left(1-
    \frac{2\cs+f\phi_b^2}{2f\phi_b^2}\ys\right).
 \label{cmeles}
\end{equation}
where factors depending on the value of $h$, the scaling function
for the potential, and its derivative evaluated at the peak
position are omitted for clarity. The above equation is in
agreement with previous results~\cite{itamar1,us} calculated in
the limit of no added-salt and negligible effect of counterions.
As the amount of salt increases, the monomer concentration
characterized by $\cm$ decreases, and for large enough amount of
added salt, the peak monomer concentration decreases below its
bulk value ---  a clear sign of depletion.

The amount of adsorbed monomers in the adsorption layer
is now calculated as function of ionic strength

\begin{eqnarray}
  \Gamma\simeq
  \cmz D_0
    \bigg[1-\frac{2\cs+f\phi_b^2}{4f}\ys
\left(\frac{1}{\phi_b^2}+\frac{2}{\cmz}\right)\bigg].
 \label{gammasalt}
\end{eqnarray}
where the added subscript zero denotes the known no-salt limits,
$\cmz=3\ys^2/(4\pi l_Ba^2)$ and $D_0=a/\sqrt{6f\ys}$, of the
maximal monomer concentration $\cm$ and adsorption length $D$,
respectively \cite{us,itamar1}.
 For no added salt, the adsorbed amount scales like
 \begin{equation}
  \Gamma=\Gamma_0\simeq  \cmz D_0\sim\ys^{3/2}f^{-1/2}l_B^{-1}a^{-1}
 \label{gammaD}
\end{equation}
When salt is added, the surface potential screening is obtained
via the PEs and  salt ions. In addition, the adsorbed amount
$\Gamma$ decreases as can be seen from the negative correction
term in Eq. (\ref{gammasalt}).

Using the above assumptions, the scaling of the induced surface
charge is:

\begin{equation}
  \left|\sigma\right|=(4\pi l_B)^{-1} \left.\frac{\dv \yy}{\dv
 x}\right|_{x=0}\sim \frac{\ys}{4\pi l_B
 D}\sim\frac{\ys^{3/2}f^{1/2}}{l_B a}
 \label{sigmagamma}
\end{equation}
Comparing this scaling to Eq.~(\ref{gammasalt}), we see that the
scaling of $\sigma$ resembles that of the charge carried by the
adsorbed amount $f\Gamma$. Subtracting the two equations shows
that the overcharging $\Delta\sigma=f\Gamma-\sigma$ scales like
$\sigma$ as well.

The numerically calculated salt dependence of the adsorption
length $D$ is presented in Fig.~4a. The location of the
concentration peak, taken as $D$, is shown as a function of the
bulk salt concentration $\cs$. The length $D$ is shown to increase
with the addition of salt, in agreement with Eq.~(\ref{Dele}) and
with experimental results~\cite{shubin,regine}. The salt
dependence of the overall adsorbed amount
$\Gamma\equiv\phi_b^2\int_0^\infty\left(\eta^2-1\right)\dv x$ is
presented in Fig.~4b. The adsorbed amount is shown to decrease
steadily with the addition of salt, in agreement with Eqs.
(\ref{gammasalt}). The sharp drop in $\Gamma$ is a sign of PE
depletion in high salt conditions, and shows that the higher terms
in the $\Gamma(\cs)$ expansion are negative as well.

We close this section by mentioning that a more elaborated
treatment of the overcharging for strongly repulsive surfaces is
presented in Appendix A. We find two different scaling regimes.
One for electrostatically dominated overcharging and the second
for excluded volume dominated one. For the electrostatically dominated
regime the overcharging follows the same scaling as the overall PE
adsorption $\Delta\sigma\sim\Gamma\sim f\sigma$. In the excluded
volume dominated regime its scaling depends on the excluded volume parameter
and yields $\Delta\sigma_0\simeq f\left|\sigma\right|
\sqrt{l_Ba^2/ (v^2\phi_b^2)}$. In both cases the magnitude of
overcharging is very small as compared to the bare $\sigma$. This
weak overcharging is not sufficient to explain PE multilayer
formation for the repulsive surfaces considered in this section.
Note that a very different situation exists when the surfaces are
chemically attractive as discussed in Sec.~\ref{phimax}.

\section{Weak Chemically-Repulsive Surfaces}
\label{append2}

In Sec.~\ref{ocreg}, we treated the strong chemically repelling
surface, and concluded that the PE-surface electrostatic
attraction is the sole drive for the adsorption. In this section,
we relax the assumption of strong repulsive surfaces, while
preserving the dominance of the electrostatic interactions.

The chemical interactions are added into our model via the amount
of PE in direct contact with the adsorbing surface
$\phi(x=0)\equiv\phi_s$ [or in the renormalized form
$\eta(x=0)=\eta_s$]. We note that the case of weakly repulsive
surfaces the slope of the monomer order parameter near the surface
must be positive $\left.\dv \eta/\dv x\right|_{x=0}>0$, as
explained in  Sec.~\ref{mfeq}.

We begin by noting that the mean field equations (\ref{normPBmod})
and (\ref{normEdwardsmod}) are invariant to translations in the
coordinate $x$. Namely, the equations are invariant under a
translational  transformation $x\rightarrow x+l$, as long as the
boundary conditions are also transformed in the same manner, {\it
i.e.} $\phi(l)=\phi_s$ and $\yy(l)=-\ys$. We note that this is a
property of the planar geometry used in this case, and that for
different geometries such as spherical or cylindrical this shift
symmetry is no longer present.

Using the above symmetry, the adsorption profile characterized by
$\phi(x),\, \yy(x)$, with boundary conditions $\phi(0)=\phi_s, \,
\yy(0)=-\ys$, can be thought of as part of a larger profile,
satisfying $\yy(-l)=-\left|\yy_s^*\right|$ and $\phi(-l)=0$. Such
a profile, in turn, is exactly similar to the one discussed in the
previous subsection, since the two systems are connected by the
above mentioned translational transformation. Therefore, finding
the above $\yy_s^*$ and $l$ from the given $\phi_s,\, \ys$ then
enables the derivation of the adsorption characteristics in the
same manner as discussed in Sec.~\ref{ocreg}.

This  strategy is demonstrated in Fig.~5. In Fig.~5a the monomer
concentration profiles are presented as a function of the distance
from the surface $x$ for several surface parameters
$0<\phi_s<16.2\phi_b$ and $0<\ys<1.0$. The profiles were obtained
by solving Eqs.~(\ref{normPBmod}) and (\ref{normEdwardsmod})
numerically using the boundary conditions $\eta(0)=\phi_s/\phi_b$,
$\yy(0)=-\ys$ close to the surface and
$\yy(x\rightarrow\infty)=0$, $\eta(x\rightarrow\infty)=1$ at
infinity. In Fig.~5b the same profiles are shown, with a
translation in the surface position. As can be seen in Fig.~5b,
all three profiles collapse on exactly the same profile, showing
that all three profiles have the same concentration and length
scales despite the difference in $\ys$ and $\eta_s$. These scales
can be calculated by using Eqs.~(\ref{Dele}), (\ref{cmeles}), as
discussed in the previous subsection.

Using the scaling relations Eqs.~(\ref{Dele}) and (\ref{cmeles}),
the above symmetry yields the following connections between
$|\yy_s^*|,\, l$ and the surface boundary conditions $\ys$ and
$\phi_s$:

\begin{eqnarray}
  \phi_s=\phi_M^* g\left(\frac{l}{D^*}\right)
 \label{phiseq}\\
  \yy_s=\left|\yy_s^*\right| h\left(\frac{l}{D^*}\right)
 \label{ys1eq}
\end{eqnarray}
where $D^*,\phi_M^*$ are the same as Eqs.~(\ref{Dele}) and
(\ref{cmeles}) when we change
$\ys\rightarrow\left|\yy_s^*\right|$. Manipulation of
Eq.~(\ref{ys1eq}) then yields:

\begin{equation}
  |\yy_s^*|=\frac{\yy_s}{h\left(l/D^*\right)}=
    \frac{\yy_s}{h\left(g^{-1}\left(\phi_s/\phi_M^*\right)\right)}
 \label{ys12eq}
\end{equation}
Eq.~(\ref{ys12eq}) can be solved iteratively for $|\yy_s^*|$
potential as a function of the surface and bulk solution
parameters. Insertion of the result into Eq.~(\ref{ys1eq}) then
gives $l/D^*$.

It is important to note, that the inversion of Eq.~(\ref{ys1eq})
is only possible in the region where $\phi$ and $x$ are monotonic,
{\it i.e.} when the surface monomer concentration is lower than
the maximal monomer concentration on the profile
$\phi_s^2<\phi_{\mathrm{max}}^2$. This condition can also be
expressed by the condition $\left.\dv \eta / \dv
x\right|_{x=0}>0$, showing that the chemical interactions between
the surface and the PE chains must still be repulsive for the
shift strategy to be employed.

We now proceed to solving Eq.~(\ref{ys12eq}) for low values of
$\phi_s/\phi_M$. Using the boundary conditions and
Eqs.~(\ref{normPBmod}, \ref{normEdwardsmod}) the characteristic
functions $h(x'),\, g(x')$ can be expanded in powers of $x'$ to
yield $g(x')=a_0x'+b_0x'^3+\dots$ and
$h(x')=-1+a_1x'+b_1x'^2+\dots$. Inserting these expansions into
Eq.~(\ref{ys12eq}) and solving to second order in $\phi_s$ yields:

\begin{equation}
  \ys^*=\ys\left[1+\frac{a_1}{a_0}\frac{\phi_s}{\phi_M}\right]+\dots
 \label{ysstareq}
\end{equation}
Note that  $\phi_M^2$ is now no longer the actual maximal monomer
concentration, but is defined in Eq.~(\ref{cmeles}) as the
characteristic for the monomer concentration, where the real
surface electrostatic potential $\ys$ is used rather than the
phantom surface potential.

We can now turn to calculate the overall adsorption. Returning to
the initial assumption that the monomer concentration profile is a
part of a larger profile starting at $x=-l$, the adsorbed amount
can be taken as the total adsorbed amount from $x=-l$ to infinity,
minus the amount adsorbed from $x=-l$ to the actual surface at
$x=0$:

\begin{eqnarray}
  \Gamma=\int_0^\infty \dv x \left(\phi^2-\phi_b^2\right) =
  \int_{-l}^\infty \dv x \left(\phi^2-\phi_b^2\right) - \nonumber
  \\
  \int_{-l}^0 \dv x \left(\phi^2-\phi_b^2\right).
 \label{gamma1eq}
\end{eqnarray}
The first integral in the RHS of Eq.~(\ref{gamma1eq}) is the
overall adsorption of PE to the phantom surface at $x=-l$, while
the second integral is the PE adsorbed amount between this phantom
surface and the real surface at $x=0$. Using the expansion of
$g(x)$ and Eq.~(\ref{phiseq}), we can see that the second integral
is of third order in $\phi_s / \phi_M$, and can be neglected for
low enough $\phi_s/\phi_M$. Using Eqs.~(\ref{gammasalt}) and
(\ref{phiseq}), Eq.~(\ref{gamma1eq}) can be evaluated as:

\begin{eqnarray}
  \Gamma\simeq \frac{\sqrt{3}\ys^{3/2}}{4\pi\sqrt{2}l_B a f^{1/2}}
  \left[1-\frac{2\cs+f\phi_b^2}{4f\phi_b^2}\ys+\right.
   \nonumber \\
  \left.\frac{3a_1\phi_s}{2a_0\phi_{M,0}}
  \left(1+\frac{2\cs+f\phi_b^2}{12f\phi_b^2}\ys\right)\right].
 \label{gammaphi0eq}
\end{eqnarray}
As expected, the adsorbed amount increases with $\phi_s$. However,
the salt dependence of the surface is very different from the
strongly repulsive surface case. For low amounts of added salt and
high enough $\phi_s$, the term combining the small ions and
$\phi_s$ is stronger than the salt term, and the adsorbed amount
{\it increases} with salt. When $\phi_s$ is low, the addition of
salt causes the adsorbed amount to decrease, similar to the
strongly repulsive surface case. In both cases, when the amount of
salt increases to a high enough value, higher order terms become
dominant, and the adsorbed amount {\it decreases}, similar to the
infinitely repulsive case in Fig.~4b.

The salt dependence of the adsorbed amount, $\Gamma$, is shown for
several $\phi_s$ values in Fig.~6a. The adsorbed amount is shown
to always increase with the amount of monomers attached to the
surface $\phi_s^2$. For low $\phi_s$ values, the adsorbed amount
decreases with salt, as expected from Eq.~(\ref{gammaphi0eq}) and
in agreement with Fig.~4b. For higher $\phi_s$ values, the
adsorbed amount is seen to increase for low bulk concentrations of
salt and then decrease strongly. In Fig.~6b the relative
overcharging is plotted for the same $\phi_s$ values, showing that
the relative overcharging is still a very small effect, even for
weakly repulsive surfaces. This shows that the attractive chemical
interactions between the surface and the PE chains are indeed
crucial for the multilayer formation, and not the electrostatic
interactions alone.


\section{Discussion and Conclusions}
\label{discussion}

%

We have presented analytical and numerical calculations of the
mean-field equations describing adsorption of polyelectrolytes
onto charged surfaces. Three surface situations are discussed:
chemically attractive surfaces, and strong and weak chemical
repulsive ones.

The strongest adsorption phenomenon is seen from numerical
solution of the mean-field equations for chemically attractive
surfaces. This manifests itself in a large overcharging of about
40\% to 55\% of the bare surface charge for high salt conditions.
On the other hand we did not find a full charge inversion as was
predicted earlier in Ref.~\cite{joanny}. This means that any model
\cite{castelnovo,castelnovo2} which tries to explain multilayer
formation at equilibrium needs to rely on non-electrostatic
complexation between the cationic and anionic polymer chains,
beside the electrostatic interactions. So far no simple scaling
form are obtained for the PE adsorbed amount in this case. However
the adsorbed amount (and the overcharging) are shown to increase
with the salt amount $\cs$, and to decrease with $f$, the charge
fraction on the PE chain as well as with $d$, which is inversely
proportional to the chemical interaction of the surface.

For the case of strong chemical repulsion between the surface and
the PE chains, we find a difference in the effect of salt addition
on the width of the adsorbed layer $D$ and on the adsorbed PE
amount $\Gamma$ (Fig.~4). The width of the PE adsorbed layer
increases with addition of salt, while the overall adsorbed PE
amount {\it decreases} with addition of salt. This difference
results from a strong decrease in the monomer concentration close
to the surface upon addition of salt. This difference between
$\Gamma$ and $D$ is in agreement with experimental results of
Shubin et al \cite{shubin}, where silica surfaces were used to
adsorb cationic polyacrylamide (CPAM). When salt is further added
to the solution, the PEs stop overcompensating the charged
surface, and we are in an under-compensation regime of adsorption.
Previous studies~\cite{us} showed that for even higher amounts of
salt, scaling as $\cs\sim f|\phi_s|$, the PEs deplete from the
charged surface. For such surfaces, the overcharging in most cases
is found to scale like the induced surface charge density
$\sigma$. For weakly charged PEs the overcharging depends on the
excluded-volume parameter and bulk monomer concentration, and
scales as $\sim f\sigma$. Very weakly charged PEs are shown to
adsorb to the charged surface, but do not overcharge it. Our
scaling results are in agreement with numerical calculations of
the mean-field equations. For all PE charges, the overcharging of
the PE with respect to a repelling surface is found to be very
small, of the order of $1\%$ of the bare surface charge. This
naturally leads to the conclusion that the overcharging relies
heavily on the chemical interactions between the surface and the
PE chains. It is of much smaller importance for this type of
repulsive surfaces than for attractive ones considered above.

For weakly repulsive surfaces, we find that the adsorbed amount
increases with the addition of salt for low salt amounts, and
decreases for high amounts of added salt. This is a natural
interpolation between the attractive and repulsive surface limits.
Moreover, the overcharging of the surface charge remains low, and
for high amounts of added salt the PE again undercompensates the
surface charges.

Our results can serve as a starting point for a more quantitative
analysis of the overcharging phenomena, and provide for better
understanding of multilayer formation.

\vspace{2truecm}

{\it Acknowledgments:} We thank I. Borukhov, Y. Burak, M.
Castelnovo, J.-F. Joanny and E. Katzav for useful discussions and
comments. In particular, we are indebted to Qiang (David) Wang for
his critical comments on a previous version of this manuscript.
Support from the Israel Science Foundation (ISF) under grant no.
210/01 and the US-Israel Binational Foundation (BSF) under grant
no. 287/02 is gratefully acknowledged.

\appendix

\section{Scaling of the Overcharging Layer in Chemically
    Repulsive Surfaces}
\label{append1}

In order to derive scaling estimates for PE adsorption regime,
the adsorbed PE layer is divided into two sub-layers around the
potential peak point $x\equiv\xfc$. The compensation layer is
defined as $x<\xfc$, and consists of PEs attracted to the surface
mainly by electrostatics. In contrast, in the overcharging
layer $(x>\xfc)$ the PE chains are electrostatically repelled
from the surface, but remain in the surface vicinity solely
because of their chain connectivity. For low enough salt
concentrations, the amount of charge carried by the PEs in the
adsorbed layer is much larger than that carried by the small
ions. In this case, the overcharging can be taken as:

\begin{eqnarray}
  \Delta\sigma\equiv  -|\sigma|+
    f\phi_b^2\int_0^\infty\dv x \left(\eta^2-1\right)
   \simeq  \nonumber \\
   f\phi_b^2\int_{x_c}^\infty\dv x \left(\eta^2-1\right),
 \label{delsig_app}
\end{eqnarray}
Where the first integral is the same as Eq.~(\ref{delsig0}), with
$\sigma$ being the {\it induced} surface charge on the adsorbing
surface.  The second integral in the above equation is taken from
$x=x_c$ to infinity, and describes the PE adsorption in the
region where the electrostatic interaction between the PE and the
surface is repulsive. The integral from $x=0$ to $x_c$ balances
the surface charge almost entirely for low amounts of added salt,
using the fact that $\left.\dv \yy /\dv x\right|_{x_c}=0$ (Gauss
law).

In order to find the amount of polymers in the overcharging
layer, we begin by examining Eq.~(\ref{PressureEq}) at the
potential peak, $x=x_c$, where the first term on the LHS
vanishes. Expanding the RHS to second order in $\yy$ without
neglecting the excluded-volume term yields:

\begin{eqnarray}
    \left.\frac{a^2}{6}\left(\frac{\dv\eta}{\dv x}\right)^2\right|_{x=\xfc}
    = f\yfc \left(\pfc^2-1\right)+ \nonumber\\
     \half \iv
    \left(\pfc^2-1\right)^2-
    \frac{f}{2}\left(\frac{\kappa^2}{k_m^2}+1\right)
        \yfc^2
 \label{fceq1}
\end{eqnarray}
where the values at the peak are denoted by $\yfc\equiv\yy(\xfc)$
and $\pfc\equiv\eta(\xfc)$. As shown in the Appendix, $\eta_c$
decreases with addition of salt. Therefore, for a large amount of
added salt, the LHS of Eq.~(\ref{fceq1}) becomes negative (this
is true to all orders of $\yfc$ and not only to order $\yfc^2$ as
shown here), while the RHS is always positive, meaning that there
is no peak in the rescaled potential $\yy$. This demonstrates
that surface charge overcharging can only occur in low enough
salt conditions.

In the overcharging layer, the previous assumptions made in
Sec. III.A about the dominance of the electrostatic interactions
are not necessarily true. Consequently, the decay of the PE concentration in
this region can be governed  by either one of two interactions:
the electrostatic or the {\it excluded-volume} repulsion between
the monomers. We consider them as
two limits for overcharging. i) An electrostatically dominated
regime, $\iv\left(\pfc^2-1\right)\ll f\yfc$,  where the excluded-volume
term is neglected in Eq.~(\ref{fceq1}). ii) The excluded-volume
dominated regime, $\iv\left(\pfc^2-1\right)\gg f\yfc$. Here, the
electrostatic interaction between monomers in Eq.~(\ref{fceq1})
is neglected. In both regimes the  value of ${\dv\eta}/{\dv
x}$ at the peak position is estimated from the scaling of the
compensation layer to be $\left.{\dv\eta}/{\dv
x}\right|_{x=\xfc}\simeq {\phi_M}/{( \phi_b D)}$.

\subsection{The strongly charged regime:
$\iv\left(\pfc^2-1\right)\ll f\yfc$} \label{elecreg}

The validity criterion of this regime is similar to the
assumptions presented in Sec.~\ref{ocreg}.A, showing that the
overcharging layer ($x>x_c$) can be thought of as an extension
of the compensation layer ($0<x<x_c$). Instead of rederiving
$\Delta\sigma$ from Eq.~(\ref{fceq1}), we can use Eq.
(\ref{pressele1}) and the expressions for $D$ and $\cm$ from
Eqs.~(\ref{Dele}) and (\ref{cmeles}), which  yield $\yfc\sim\ys$
and $\pfc^2\sim\cm/\phi_b^2$. The resulting overcharging scales
like the bare surface charge:

\begin{equation}
  \Delta\sigma\sim f\Gamma_D \sim f(\cm-\phi_b^2)D\sim\left|\sigma\right|.
 \label{delsigele}
\end{equation}
Note that the effect of added salt in this regime is similar to
the one described  in Eq. (\ref{gammasalt}). Namely, the added
salt lowers the surface charge overcharging, in agreement with
experimental~\cite{shubin} and numerical~\cite{us} results.

\subsection{The intermediate charged regime:
$\iv\left(\pfc^2-1\right)\gg f\yfc$ } \label{xvreg}

In this regime, the excluded-volume interactions dominate the
decay of the PE concentration. Using the regime validity condition
and neglecting the first term in the RHS of Eq.~(\ref{fceq1})
yields an expression for the monomer concentration at $\xfc$:

\begin{equation}
   \pfc^2=1+\sqrt{\frac{a^2}{3\iv}\left.\left(\frac{\dv \eta}{\dv
   x}\right)^2\right|_{x=\xfc}+\frac{\kappa^2+k_m^2}{k_m^2\iv}f\yfc^2}
 \label{pfceqxv1}
\end{equation}
Noting that the ${\dv\eta}/{\dv x}$ is continuous at $x=\xfc$, we
use the compensation layer scaling estimates to find
$\left.{\dv\eta}/{\dv x}\right|_{x=\xfc}\simeq \phi_M / (\phi_b
D)$. Substituting the latter into Eq.~(\ref{pfceqxv1}) and expanding to
first order in both the ionic strength
$2\cs+f\phi_b^2=(\kappa^2+k_m^2)/4\pi l_B$ and the ratio of the
bulk and peak monomer concentrations $\phi^2_b/\phi_M^2\simeq
l_Ba^2\phi_b^2/\ys^2 $ yields:

\begin{eqnarray}
  \pfc^2\simeq 1+ \sqrt{\frac{3}{2\pi}}\frac{\ys^{3/2}f^{1/2}}
    {\sqrt{ l_B a^2 v}\phi_b^2}
    \bigg[1-\frac{2\cs+f\phi_b^2}{2f\phi_b^2}\ys+ \nonumber \\
    \frac{2\pi l_Ba^2\phi_b^2}{3\ys^2}\bigg].
 \label{pfceqxv}
\end{eqnarray}
where Eq.~(\ref{Dele}) and (\ref{cmeles}) are used for $D$ and
$\cm$, respectively.

The overcharging is calculated from Eq.~(\ref{delsig0}) where the
characteristic length scale entering the integral is the Edwards
length, $\xi_e=a/\sqrt{3\iv}$, depending only on the
excluded-volume interactions and not on the salt. The overcharging
$\Delta\sigma \simeq f\phi_b^2\left(\pfc^2-1\right) \xi_e$ is:

\begin{eqnarray}
  \Delta\sigma \simeq
    \frac{\ys^{3/2}f^{3/2}}{\sqrt{2\pi l_B\phi_b^2} v}
    \bigg[1-\frac{2\cs+f\phi_b^2}{2f\phi_b^2}\ys+  \nonumber\\
     \frac{2\pi l_Ba^2\phi_b^2}{3\ys^2}\bigg]
 \label{delgamxv}
\end{eqnarray}
In the limit of no ionic strength ($\cs=0$ and negligible
counterion contribution), the overcharging from Eq.
(\ref{delgamxv}) scales like $\Delta\sigma_0\simeq f\left|\sigma\right|
\sqrt{l_Ba^2/ (v^2\phi_b^2)}$. Addition of salt results in a
decrease of overcharging, similar to what was shown in the previous
subsection for the strongly charged regime.

By comparing the corresponding expressions for $f\yfc$ and
$\iv(\pfc^2-1)$ in both regimes, we conclude  that the boundary
between the strongly charged and intermediate regimes occurs at
$f\simeq 3v \ys/(2\pi l_B a^2)$ (see also Ref. \cite{itamar1}).
This serves as a self consistency test.

We define  the adsorption excess
$\Delta\Gamma$ by integrating
numerically the adsorbed amount from $x_c$ to $\infty$. Note that in the low salt limit,
$f\Delta\Gamma\simeq \Delta \sigma$,  can be thought of as the
overcharging parameter used earlier.
It is
then possible to make a direct comparison with the two limits
discussed above. The adsorption excess as a function of $f$ is
presented in Fig.~7a. For small $f$ values, $\Delta\Gamma$ scales
like $f^{1/2}$, in agreement with Eq.~(\ref{delgamxv}), while for
larger values of $f$ the excess adsorption $\Delta\Gamma$ scales
like $f^{-1/2}$, in agreement with Eqs.~ (\ref{gammaD}) and
(\ref{delsigele}). These two limiting scaling behaviors are shown
on Fig.~7a. In Fig.~7b, $\Delta\Gamma$ is presented as a function
of $\ys$. The scaling $\Delta\Gamma \sim\ys^{3/2}$ also shown in
the figure, is in agreement with Eqs.~(\ref{gammaD}), (\ref{delsigele})
and (\ref{delgamxv}). We note, by comparing
Fig.~4b to Fig.~7, that the overcompensating PE are of the order
of less than $1\%$ of the total adsorbed amount, showing that the
mean-field overcharging of repulsive surfaces is an extremely
small effect.


\subsection{The undercompensation threshold}
\label{app2}

So far we presented the case where the PE layer is
overcompensating the surface charge, and discussed it in two
limits of strong and intermediate charged PEs. In some range of
system parameters the PE charges do not overcompensate the
surface ones. The threshold for having this {\it
undercompensation} is briefly discussed here.

Re-examining the validity of Eq.~(\ref{fceq1}), it can be seen
that for high enough salt the third term on the right dominates the
largest of the first two terms when:

\begin{equation}
  \cs+\half f\phi_b^2>\frac{f\ys}{l_Ba^2}
 \label{csund}
\end{equation}
up to some numerical pre-factors. Note that
the same inequality is valid in both limits of strong and
intermediate regimes, as long as we are in high salt conditions.
It should be noted, that a similar
scaling rule was found in Refs.\cite{wiegel,us,dobrynin,muthu} for
the adsorption-depletion crossover. Numerical results show that
the overcharging-undercompensation transitions indeed have
the same scaling, but differ in the constant multiplying the
scaling result.

\vspace{2truecm}




\section*{Figure Captions}

\begin{itemize}

\item[{\bf Fig. 1}]
    The relative overcharging, $\Delta\sigma / |\sigma|=(f\Gamma-|\sigma|)/|\sigma|$,
    is presented as
    a function of the amount of added salt, $\cs$. The solid line
    corresponds to the numerical results (Sec.~\ref{phimax}), while the dashed line
    corresponds to the  predictions from Eq.~(9) in
    Ref.~\cite{joanny}. As the surface potential, $\ys$, is fixed,
    $\sigma$ is the numerically calculated
    induced surface charge.
    The numerical results show
    a much lower overcharging than the predicted ones. Furthermore,
    the salt dependence of the overcharging is shown to be much stronger.
    The system parameters have been chosen to match those of  Ref.~\cite{joanny}.
    A dilute aqueous solution ($\phi_b=0$), in theta solvent
    conditions ($v=0$). Other parameters are $\varepsilon=80$, $T=300$\,K, $\ys=1.0$,
    $a=5$\,\AA, $f=1$, $d=100$\,\AA.

\item[{\bf Fig. 2}]
    The relative overcharging, $\Delta\sigma/|\sigma|$, is plotted against
    the chemical interaction parameter $d$, for (a) $f=0.2$ and (b) $f=1$.
    The  solid line corresponds to the numerically calculated
    relative overcharging, while the dashed line corresponds to the
    theoretical predictions taken from Ref.~\cite{joanny}. Unlike the previous
    prediction, we do not observe a full charge inversion.
    The calculations are done in high salt conditions
    $\cs=1$\,M, corresponding to a Debye length of about $3$\,\AA. Other
    parameters used are the same as in Fig.~1. Although the numerical and
    predicted profiles show a similar qualitative dependence with $d$,
    they do not coincide, and converge to different
    values at $d\to\infty$. For practical purposes, $d=100$\,\AA,
    is already a very good approximation for the indifferent $d\to\infty$
    surface.

\item[{\bf Fig. 3}]
    (a) The numerically calculated adsorbed amount
    $\Gamma=\int_0^\infty\dv x\left(\phi^2-\phi_b^2\right)$ is plotted as
    a function of the amount of added salt for three values of
    the surface potential, $\ys$. The
    solid line corresponds to $\ys=1.0$, the dashed-dotted line to
    $\ys=0.5$ and the dashed line to $\ys=0.2$. The adsorbed amount is
    shown to increase both with the amount of added salt and with $\ys$.
    Other parameters used are as in Fig.~1.
    In (b) the same results are plotted for
    $\Delta\sigma/|\sigma|$
    as function of the amount of added
    salt.
    The relative overcharging $\Delta\sigma/|\sigma|$ is seen to decrease with $\ys$, in contrast
    to $\Gamma$ where it increases (see part (a)). This implies that
    the surface charge $\sigma$
    has a stronger dependence on $\ys$ than the adsorbed amount $\Gamma$.
    Therefore, the adsorbed amount
    no longer scales with $\sigma$ for attractive surfaces,
    in contrast to repulsive surfaces, Eq.~(\ref{gammaD}).

\item[{\bf Fig. 4}]
        (a) The width of the concentration profile, $D$, taken as the peak
        location, is presented as a function of, $\cs$, the added-salt
        concentration. The dotted line corresponds to $f=0.1$, the dashed line
        to $f=0.18$, the dashed-dotted line to $f=0.56$ and the solid line
        to $f=1$. Other parameters used are $\varepsilon=80$, $T=300$\,K,
        $\ys=1$, $a=5$\,\AA, $\phi_b^2=10^{-6}$\,\AA$^{-3}$, ${\it v}=50$\,\AA$^3$.
        The length scale of the adsorption $D$ is seen to increase
        with the addition of salt. For high enough added salt concentrations,
        the concentration peak vanishes altogether, indicating that the
        polymer is depleted from the surface. The adsorption-depletion
        crossover is denoted by a full circle.
    (b) The total adsorbed amount
        $\Gamma=\int_0^\infty \dv x\left(\phi^2-\phi_b^2\right)\sim c_mD$
        is plotted against the amount of added salt. The solid line
        corresponds to $f=0.03$, the dashed line to $f=0.1$, the dash-dot
        to $f=0.31$ and the dotted line to $f=1$. The adsorbed amount
        decreases slowly with salt for low amounts of added salt. For high
        concentrations of added salt the adsorbed amount decreases
        sharply to negative values, signaling an adsorption-depletion
        transition. Other parameters used are the same as in (a). (reproduced
    from Ref.~\cite{us}).

\item[{\bf Fig. 5}]
    (a) The numerically calculated monomer concentration $\eta^2(x)$
    is plotted as a function of the distance from the charged surface
    $x$, for several values of $\ys$ and $\eta_s\equiv \phi_s/\phi_b$.
    The solid line corresponds to
    $\eta_s=0$ and $\ys=1.0$, the dashed line with triangle
    markers to $\eta_s=12.55$ and $\ys=0.6$, and the dashed-dotted
    line with square markers to $\eta_s=16.2$ and $\ys=0.34$. All
    profiles share $\cs=0.01$\,M, $f=1$, $a=5$\,\AA, $v=10$\,\AA$^3$,
    $\phi_b^2=10^{-6}$\,\AA$^{-3}$, $\varepsilon=80$ and $T=300$\,K.
    All profiles are found to have the same height in the peak monomer
    concentration, despite the difference in the boundary conditions.
    (b) the same profiles as in part (a), after a shift in the
    $x=0$ position is used. The solid line is not shifted, the dashed line
    with triangular markers is shifted by $\Delta x=2.35$\,\AA\
    and the dashed-dotted line with square markers is shifted by
    $\Delta x=4.12$\,\AA. All three profiles show a data collapse on
    the solid line profile, corresponding to $|\yy_s^*|=1.0$.

\item[{\bf Fig. 6}]
    (a) Numerically calculated adsorbed amount of monomers $\Gamma$ is
    plotted against the added salt concentration $\cs$ for
    several values of $\eta_s=\phi_s/\phi_b$, for the case of weak chemically repulsive
    surfaces. The solid line corresponds
    to $\eta_s=5$, the dashed line corresponds to $\eta_s=50$ and
    the dashed-dotted line to $\eta_s=100$. All profiles share
    $\ys=1.0$, $f=1.0$, $a=5$\,\AA, $v=10$\,\AA$^3$,
    $\phi_b^2=10^{-8}$\,\AA$^{-3}$, $\varepsilon=80$ and $T=300$\,K.
    The adsorbed amount
    is seen to increase with $\eta_s$ for all values of $\cs$. For
    low amounts of added salt, the adsorbed amount is seen to increase
    with salt, in contrast to the $\eta_s=0$ case (strong repulsive surface) in Fig.~4b,
    while for high amounts of added salt the adsorbed amount
    decreases, in agreement with Fig.~4b.
    (b) The relative overcharging
    $\Delta\sigma/\left|\sigma\right|$
    is plotted against the amount of added salt for the same parameters as
    in part (a). Despite the increase in the adsorbed amount, the
    relative overcharging  remains a very small
    effect. At high salt concentration, the relative
    overcharging becomes negative --- signaling an {\it
    under-compensation} of the surface charge.

\item[{\bf Fig. 7}]
    a) Numerically calculated excess adsorption
    $\Delta\Gamma$,
    defined as the PE adsorbed amount from the potential peak to infinity,
    is plotted against $f$. The squares correspond to $\ys=0.5$,
    and the triangles to $\ys=0.6$. Both profiles share
    $\phi_b^2=10^{-6}$\,\AA$^{-3}$, $v=10^2$\,\AA$^3$, $a=5$\,\AA,
    $\cs=0.1\,$mM, $T=300$\,K and $\varepsilon=80$. The two profiles
    can be fitted in the low $f$ region by $\Delta\Gamma\sim f^{1/2}$
    (dashed line), followed by a high $f$ region where
    $\Delta\Gamma\sim f^{-1/2}$ (solid line). These
    scaling results are in agreement with Eqs.~(\ref{delsigele}) and
    (\ref{delgamxv}).
    b) $\Delta\Gamma$ is plotted against the surface potential $\ys$.
    The squares correspond to $f=0.1$,
    and the triangles to $f=0.3$. All other parameters are the same
    as in (a). The two profiles show
    a scaling of $\Delta\Gamma\sim\ys^{3/2}$, fitted by a solid line,
    in agreement with Eqs.~(\ref{delsigele}) and (\ref{delgamxv}).
    The constant prefactors in the fitting lines are
    obtained by imposing the condition that the fitting line passes through the
    last numerical data point in the respective regime.

\end{itemize}

\newpage


\noindent \textbf{Fig.~1 Shafir and Andelman}:

\vspace{1.5cm}

\scalebox{0.7}{\includegraphics{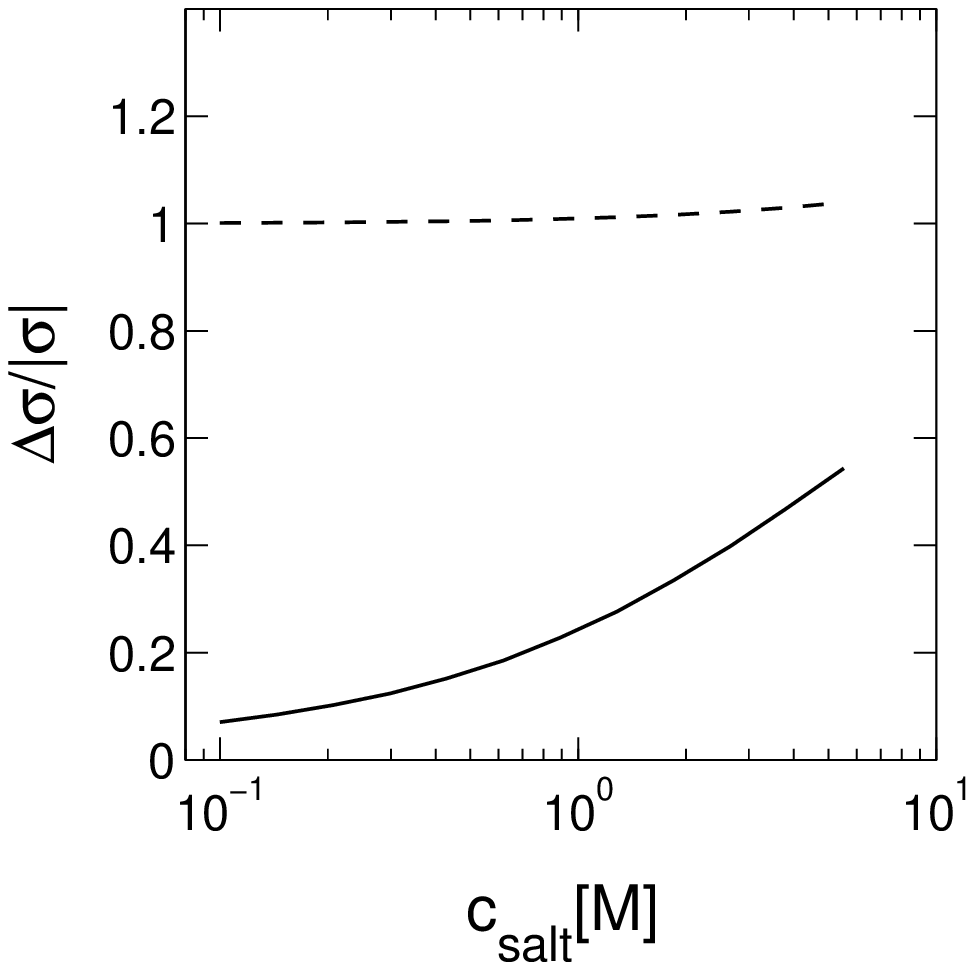}}

\vspace{3cm}

\newpage
\begin{widetext}

\noindent \textbf{Fig.~2 Shafir and Andelman}:

\vspace{1.5cm}

\scalebox{0.7}{\includegraphics{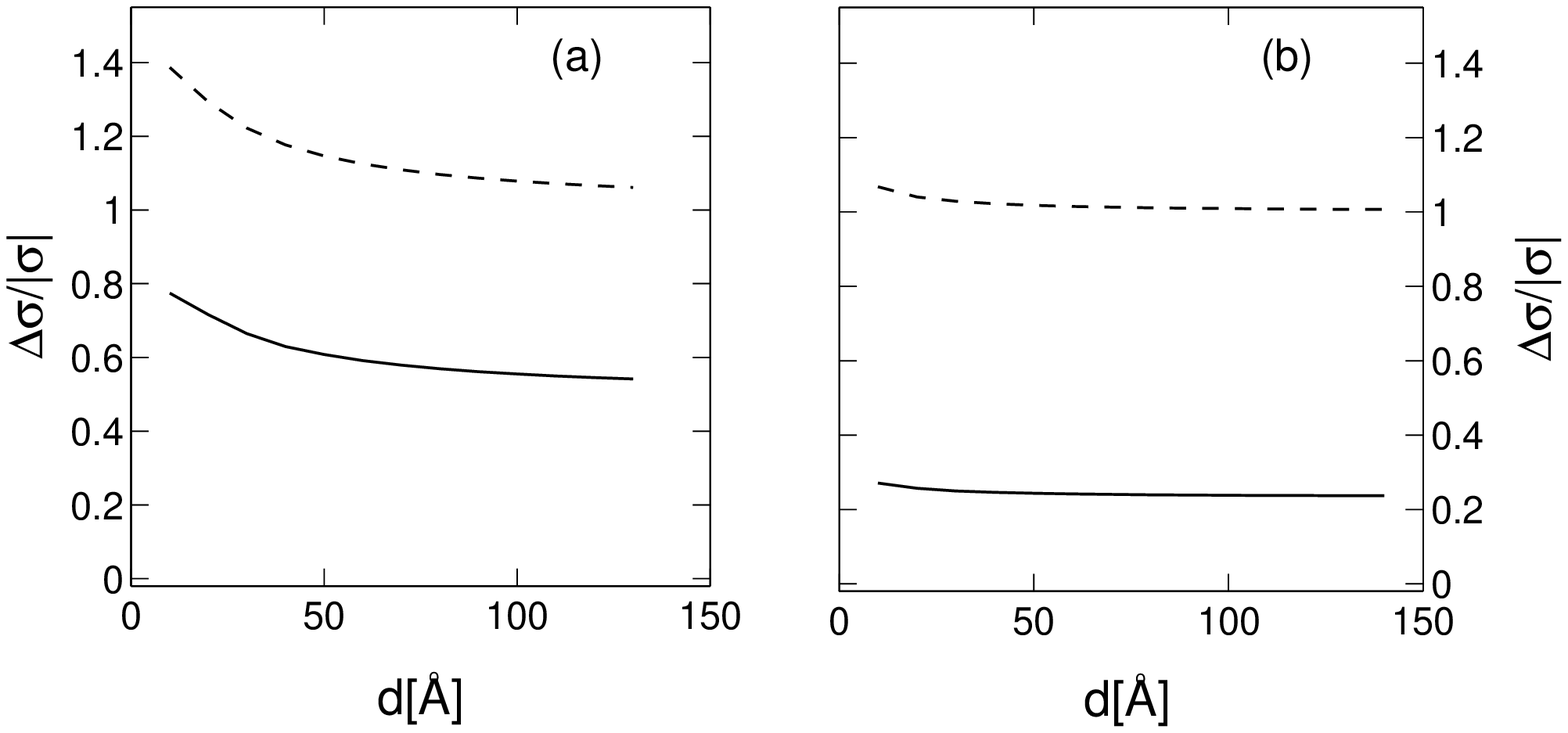}}

\vspace{3cm}

\noindent \textbf{Fig.~3 Shafir and Andelman}:

\vspace{1.5cm}

\scalebox{0.7}{\includegraphics{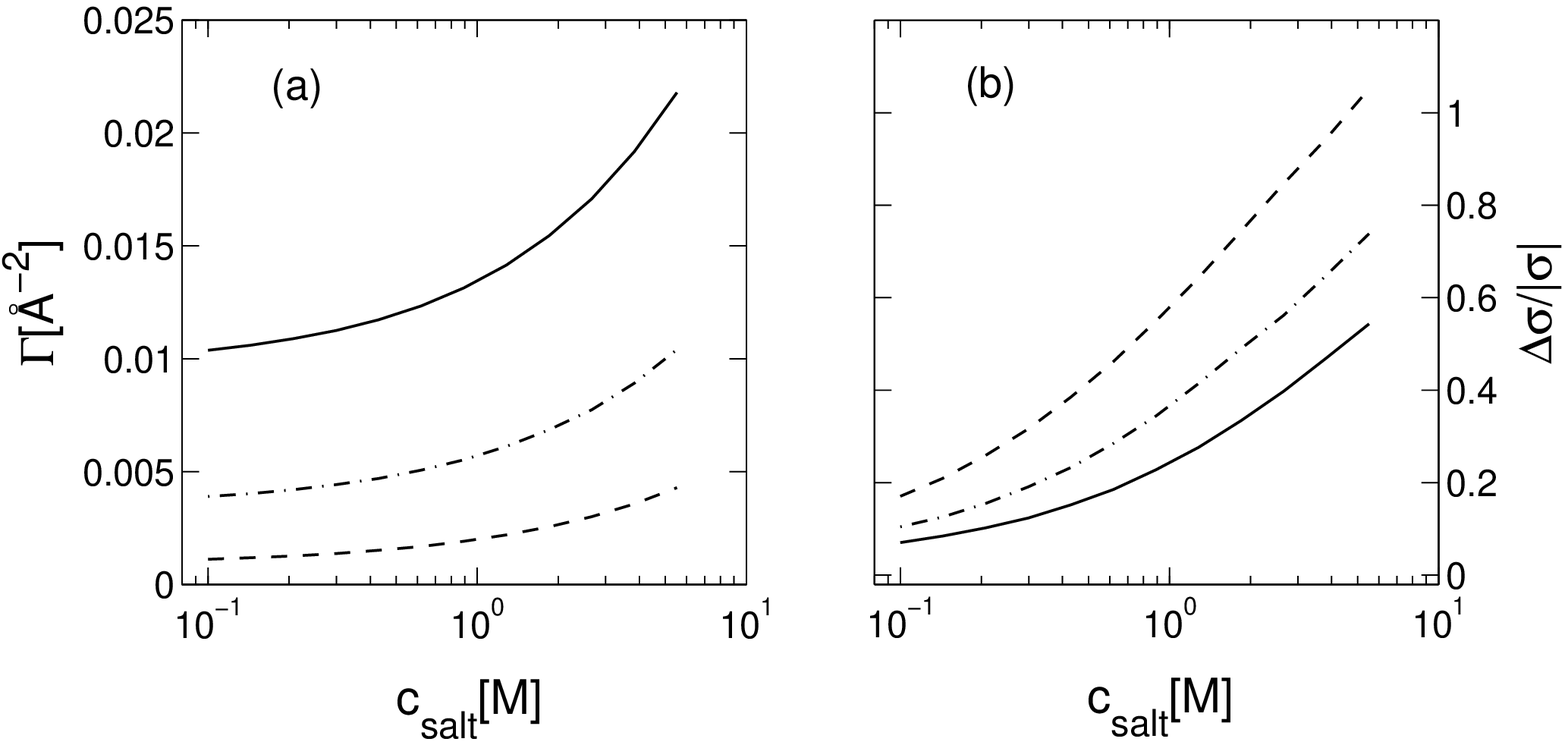}}

\vspace{3cm}

\noindent \textbf{Fig.~4 Shafir and Andelman}:

\vspace{1.5cm}

\scalebox{0.7}{\includegraphics{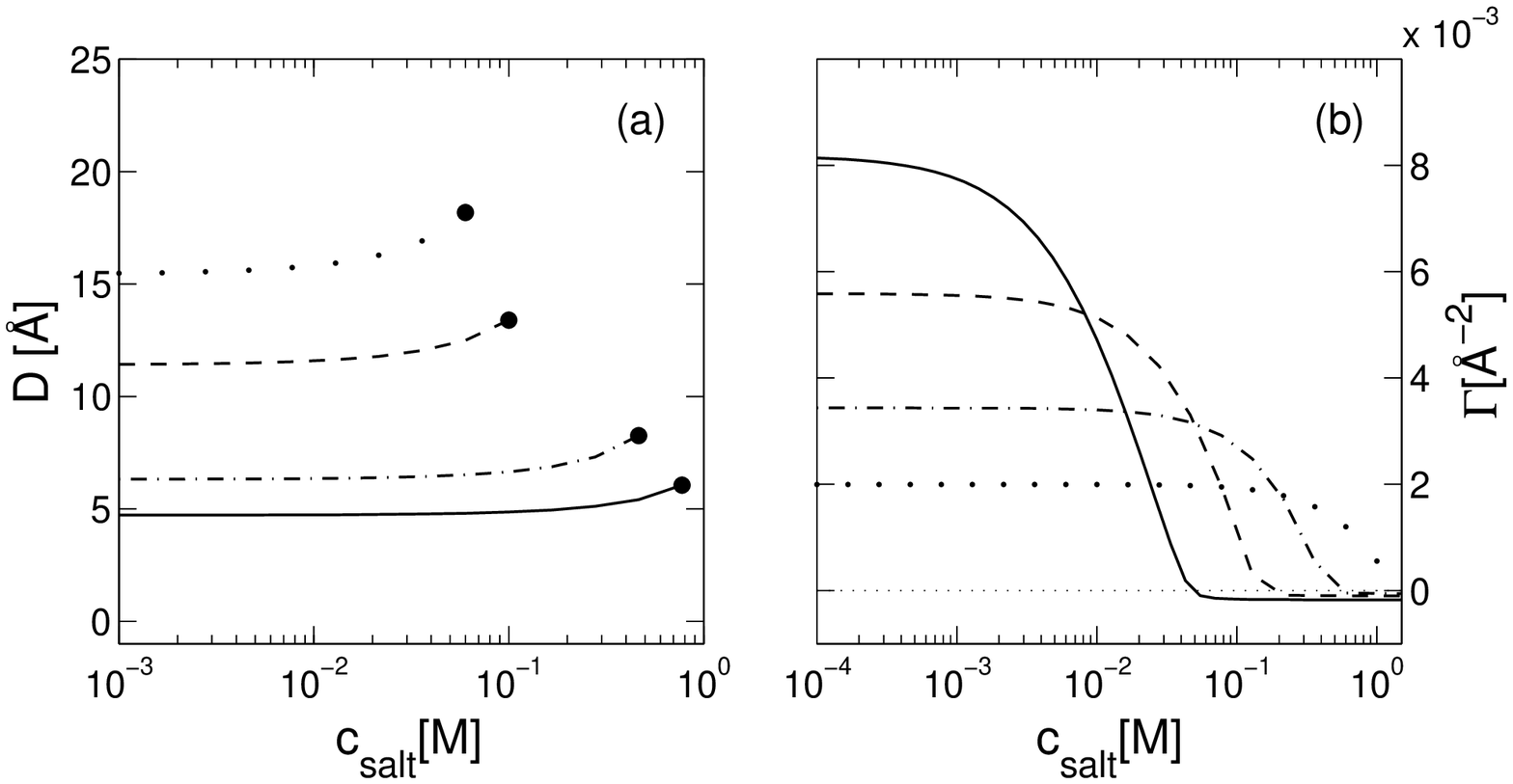}}

\vspace{3cm}

\noindent \textbf{Fig.~5 Shafir and Andelman}:

\vspace{1.5cm}

\scalebox{0.7}{\includegraphics{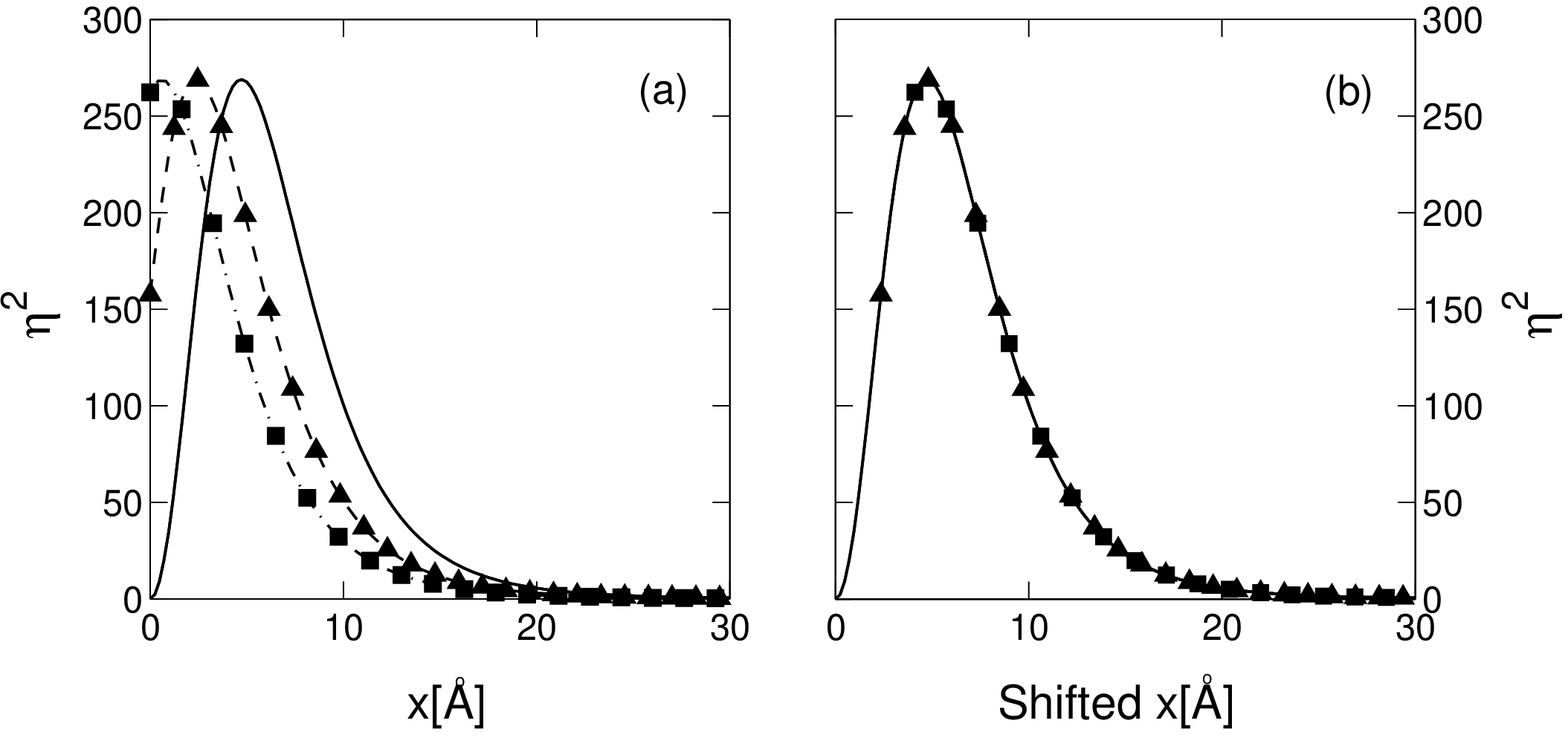}}

\vspace{3cm}

\noindent \textbf{Fig.~6 Shafir and Andelman}:

\vspace{1.5cm}

\scalebox{0.7}{\includegraphics{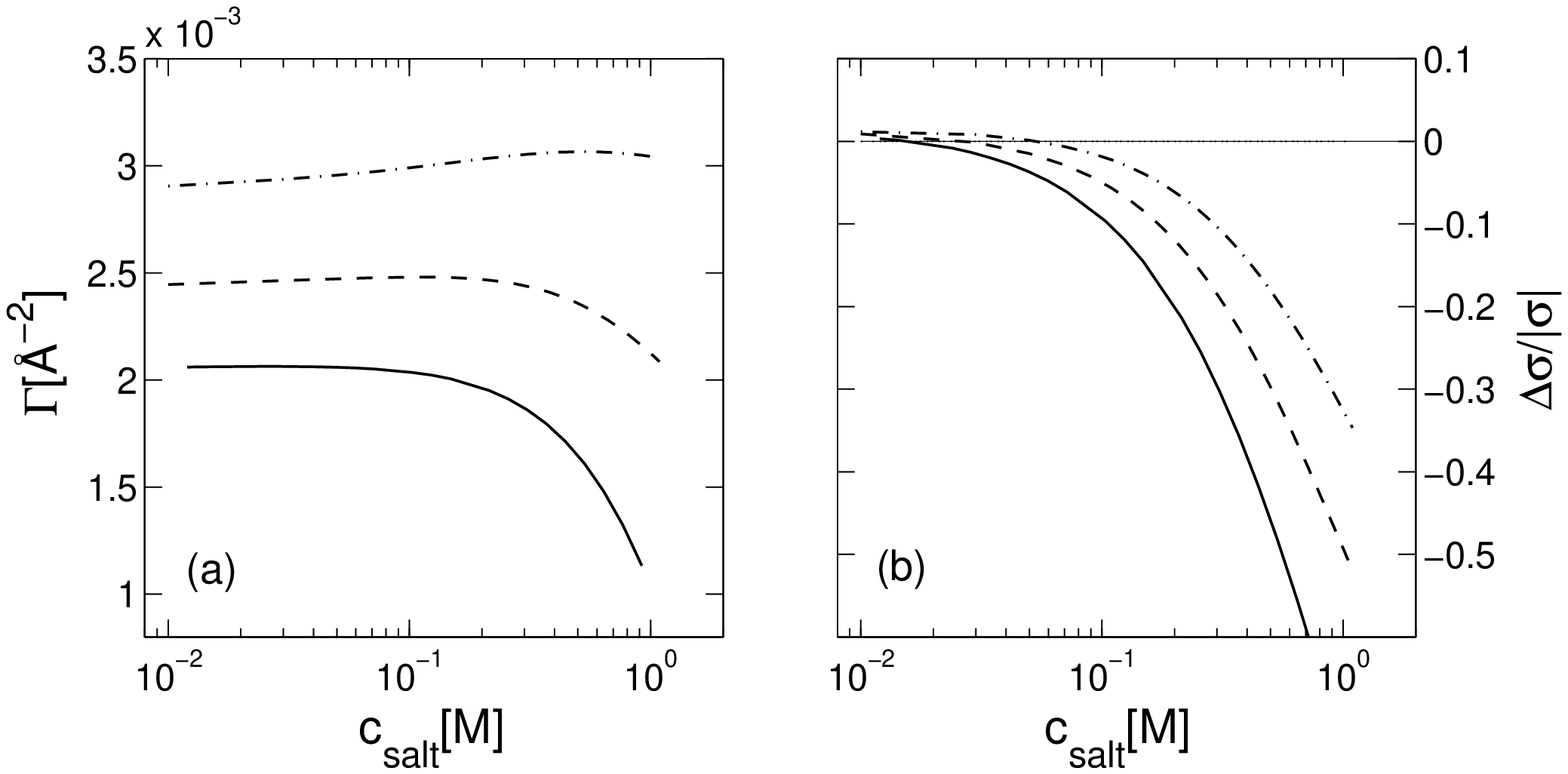}}

\vspace{3cm}

\noindent \textbf{Fig.~7 Shafir and Andelman}:

\vspace{1.5cm}

\scalebox{0.7}{\includegraphics{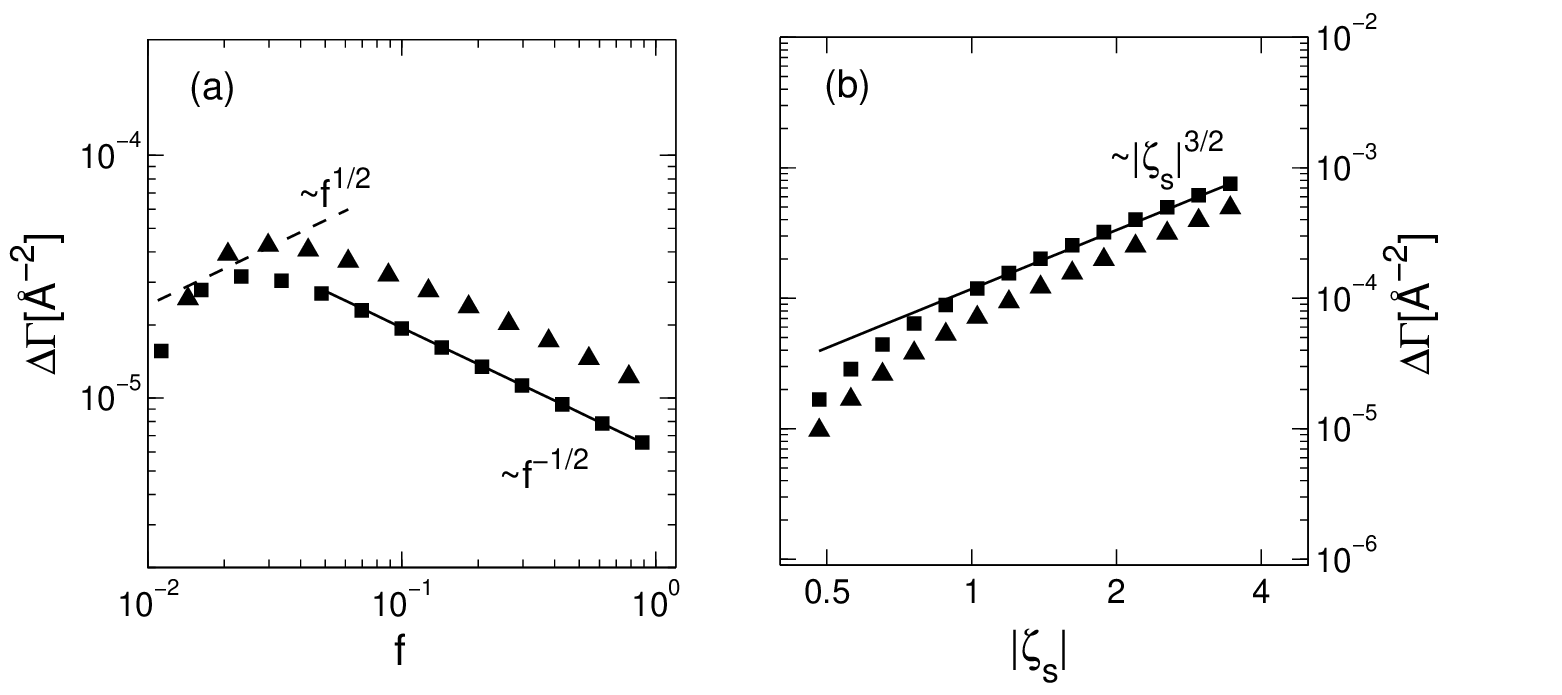}}

\vspace{3cm}

\end{widetext}


\begin{thebibliography}{99}


\bibitem{wiegel} F.W. Wiegel, J. Phys A: Math. Gen. {\bf 10},
     299 (1977).

\bibitem{muthu} M. Muthukumar, J. Chem. Phys. {\bf 86}, 7230 (1987).

\bibitem{chatellier} X. Chatellier and J.F. Joanny, J. Phys. II (France)
    {\bf 6}, 1669 (1996).

\bibitem{varoqui2} R. Varoqui, A. Johner, A. Elaissari, J. Chem. Phys.
         {\bf 94}, 6873 (1991).

\bibitem{varoqui1} R. Varoqui, J. Phys. II (France) {\bf 3}, 1097 (1993).

\bibitem{joanny} J.F. Joanny, Eur. Phys. J. B. {\bf 9}, 117 (1999).

\bibitem{itamar1} I. Borukhov, D. Andelman, H. Orland, Macromolecules
    {\bf 31}, 1665 (1998); Europhys. Lett. {\bf 32}, 499 (1995).

\bibitem{itamar2} I. Borukhov, D. Andelman, H. Orland, {Eur. Phys. J. B}
    {\bf 5}, 869 (1998).

\bibitem{itamar3} I. Borukhov, D. Andelman, H. Orland, J. Phys. Chem. B
    {\bf 103}, 5042 (1999).

\bibitem{us} A. Shafir, D. Andelman, R.R. Netz, J. Chem. Phys. {\bf 119},
    2355 (2003).

\bibitem{manghi} M. Manghi and M. Aubouy, cond-mat/0202045 preprint.

\bibitem{borisov} O.V. Borisov, E.B. Zhulina, T.M. Birshtein,
     J. Phys. II (France) {\bf 4}, 913 (1994).

\bibitem{netz} R.R. Netz and J.F. Joanny, Macromolecules {\bf 32}, 9013 (1999).

\bibitem{dobrynin} A.V. Dobrynin, A. Deshkovski, M. Rubinstein, Macromolecules
    {\bf 34}, 3421 (2001).

\bibitem{review1} R.R. Netz and D. Andelman,
         in: {\it Encyclopedia of Electrochemistry}, Eds. M. Urbakh and
    E. Giladi, Vol. I, (Wiley-VCH, Weinheim, 2002).

\bibitem{review} R.R. Netz and D. Andelman, Phys. Rep. {\bf 380}, 1 (2003).

\bibitem{vanderschee} H.A. van der Schee and J. Lyklema, J. Phys. Chem.
     {\bf 88}, 6661 (1984).

\bibitem{fleer} O.A. Evers, G.J. Fleer, J.M.H.M. Scheutjens, J. Lyklema,
    J. Coll. Interface Sci. {\bf 111}, 446 (1986).

\bibitem{bohmer} M.R. B\"{o}hmer, O.A. Evers, J.M.H.M. Scheutjens,
     Macromolecules {\bf 23}, 2288 (1990).

\bibitem{vandesteeg} H.G.M. van de Steeg, M.A. Cohen Stuart, A. de Keizer,
    B.H. Bijsterbosch, Langmuir {\bf 8}, 2538 (1992).

\bibitem{yamakov} V. Yamakov, A. Milchev, O. Borisov, B. D\"{u}nweg,
    J. Phys: Condens. Matter {\bf 11}, 9907 (1999).

\bibitem{muthu2} M. Ellis, C.Y. Kong, M. Muthukumar, J. Chem. Phys. {\bf 112},
    8723 (2000).

\bibitem{muthu3} J. McNamara, C.Y. Kong, M. Muthukumar, J. Chem. Phys.
    {\bf 117}, 5354 (2002).

\bibitem{mohwald} Yu. Lvov, G. Decher, H. Haas, H. M\"ohwald, A.
Kalachev, Physica B {\bf 198}, 89 (1994).

\bibitem{decher} G. Decher, Science {\bf 277}, 1232 (1997).


\bibitem{regine} U. Voigt, V. Khrenov, K. Tauer, M. Halm, W. Jaeger,
    R. von Klitzing, J. Phys.: Condens. Matter {\bf 15}, S213 (2003).

\bibitem{castelnovo} M. Castelnovo, J.F. Joanny, Langmuir {\bf 16}, 7524
    (2000).

\bibitem{castelnovo2} M. Castelnovo, J.F. Joanny, Eur. Phys. J. E. {\bf 6},
    377 (2001).

\bibitem{degennes2} P.G. de Gennes, Macromolecules {\bf 14}, 1637 (1981).

\bibitem{shubin} V. Shubin, P. Linse, J. Phys. Chem. {\bf 99}, 1285 (1995).






\end{thebibliography}
\end{document}